\documentclass[aps,prl,twocolumn,english,balance,superscriptaddress,floats,showpacs,letter]{revtex4-1}
\usepackage[T1]{fontenc}
\usepackage[latin9]{inputenc}
\usepackage{amsmath}
\usepackage{amssymb}
\usepackage{stmaryrd}
\usepackage{graphicx}
\usepackage{esint}
\usepackage{subfigure}
\usepackage{multirow}

\makeatletter

\newcommand{\beq}{\begin{equation}}
\newcommand{\eeq}{\end{equation}}
\newcommand{\bea}{\begin{eqnarray}}
\newcommand{\eea}{\end{eqnarray}}
\newcommand{\bwt}{\begin{widetext}}
\newcommand{\ewt}{\end{widetext}}
\@ifundefined{textcolor}{}
{%
 \definecolor{BLACK}{gray}{0}
 \definecolor{WHITE}{gray}{1}
 \definecolor{RED}{rgb}{1,0,0}
 \definecolor{GREEN}{rgb}{0,1,0}
 \definecolor{BLUE}{rgb}{0,0,1}
 \definecolor{CYAN}{cmyk}{1,0,0,0}
 \definecolor{MAGENTA}{cmyk}{0,1,0,0}
 \definecolor{YELLOW}{cmyk}{0,0,1,0}
}

\newcommand{\fvec}[1]{\boldsymbol{#1}}

\newcommand{\rmd}{{\rm d}}

\usepackage{babel}

\makeatother

\usepackage{babel}

\begin{document}

\title{Symmetry, maximally localized Wannier states, and low energy model for the twisted bilayer graphene narrow bands}

\author{Jian Kang}
\email{jian.kang@fsu.edu}
\affiliation{National High Magnetic Field Laboratory, Tallahassee, Florida, 32304, USA}

\author{Oskar Vafek}
\email{vafek@magnet.fsu.edu}
\affiliation{National High Magnetic Field Laboratory, Tallahassee, Florida, 32304, USA}
\affiliation{Department of Physics,
Florida State University, Tallahassee, Florida 32306, USA}

\begin{abstract}
 We build symmetry adapted maximally localized Wannier states, and construct the low energy tight binding model for the four narrow bands of the twisted bilayer graphene. We do so when the twist angle is commensurate, near the `magic' value, and the narrow bands are separated from the rest of the bands by energy gaps. On each layer and sublattice, every Wannier state has three peaks near the triangular Moire lattice sites.  However, each Wannier state is localized and centered around a site of the  honeycomb lattice that is dual to the triangular Moire lattice. Space group and the time reversal symmetries are realized locally. The corresponding tight binding model  provides a starting point for studying the correlated many-body phases.
\end{abstract}

\maketitle

\emph{Introduction.}
The discovery of superconductivity and correlated insulator(s) in the `magic' angle twisted bilayer graphene~\cite{Pablo1, Pablo2}  has resulted in a remarkable flurry of theoretical activity~\cite{MacDonald, Balents, Liang, Senthil, Juricic, Scalettar, Baskaran, Phillips, Kivelson, Ma, Zhang, Das, Yang, Patrick, Heikkila, Spalek}.
Central theoretical challenges are to understand the nature and the mechanism of the insulator(s) and the superconductor.
Should the most prominent insulating states -- which onset at two electrons/holes per triangular Moire unit cell, i.e. at quarter filling of the four narrow bands -- be thought of as a largely featureless Mott state in which charge motion is arrested by the Coulomb repulsion, or is a spontaneously broken symmetry responsible for the charge gap?
Is the superconductivity unconventional in that it breaks some of the lattice symmetries and perhaps originates from the electron-electron repulsion without a major role from electron-phonon interaction, or is it conventional?

In order to address the above questions, it is necessary to first construct a realistic, but simple, model of the electron motion in the narrow bands.
As pointed out in Refs.~\cite{Senthil,Liang}, this is not an obvious task.
When the twist angle is commensurate, the Moire pattern becomes periodic and leads to the triangular super-lattice, see e.g.~\cite{Mele}. At small twist angles, a unit cell contains a large number of carbon atoms and consequently the Moire Brillouin zone (MBz) becomes small. Indeed, the low energy band structure of the twisted bilayer graphene (tBG) differs in important aspects from that of two isolated monolayers due to the sizable interlayer tunneling.  The four bands around the charge neutrality point have a strongly reduced bandwidth and Fermi velocity. When the twist angle is fine-tuned to the `magic' values, the band-width becomes very narrow (but non-zero), the Fermi velocity at Dirac cones vanishes, and the quadratic band touching points appear at the corners of the MBz~\cite{Pablo1}.

Although the local charge density at quarter filing is peaked at the triangular Moire lattice sites~\cite{Pablo1}, as recognized in Refs.~\cite{Liang,Senthil} the salient features of the narrow band structure cannot be recovered unless the Wannier states (WSs) are centered at the dual honeycomb sites. We prove this using different arguments below. In addition, we diagonalize a microscopic tight binding model with the large number of atoms in the unit cell according to the prescription by Moon and Koshino~\cite{Koshino}. Based on the layer and the microscopic carbon sublattice structure of the resulting Bloch states at the MBz center, we construct the initial ansatz for the localized WSs which we project onto the Hilbert space spanned by the four narrow bands~\cite{Vanderbilt}. By construction, our ansatz realizes the lattice and the time reversal symmetries locally, and forms a \emph{non-trivial} representation of the site symmetry group. The result is then used as the initial step in the iterative procedure of Marzari and Vanderbild~\cite{Vanderbilt} to construct maximally localized, yet symmetry adapted~\cite{Sakuma}, WSs. They are then used to construct the low energy tight binding model.

\begin{figure}[htbp]
\centering
\subfigure[\label{Fig:lattice:Superlattice}]{\includegraphics[width=0.48\columnwidth]{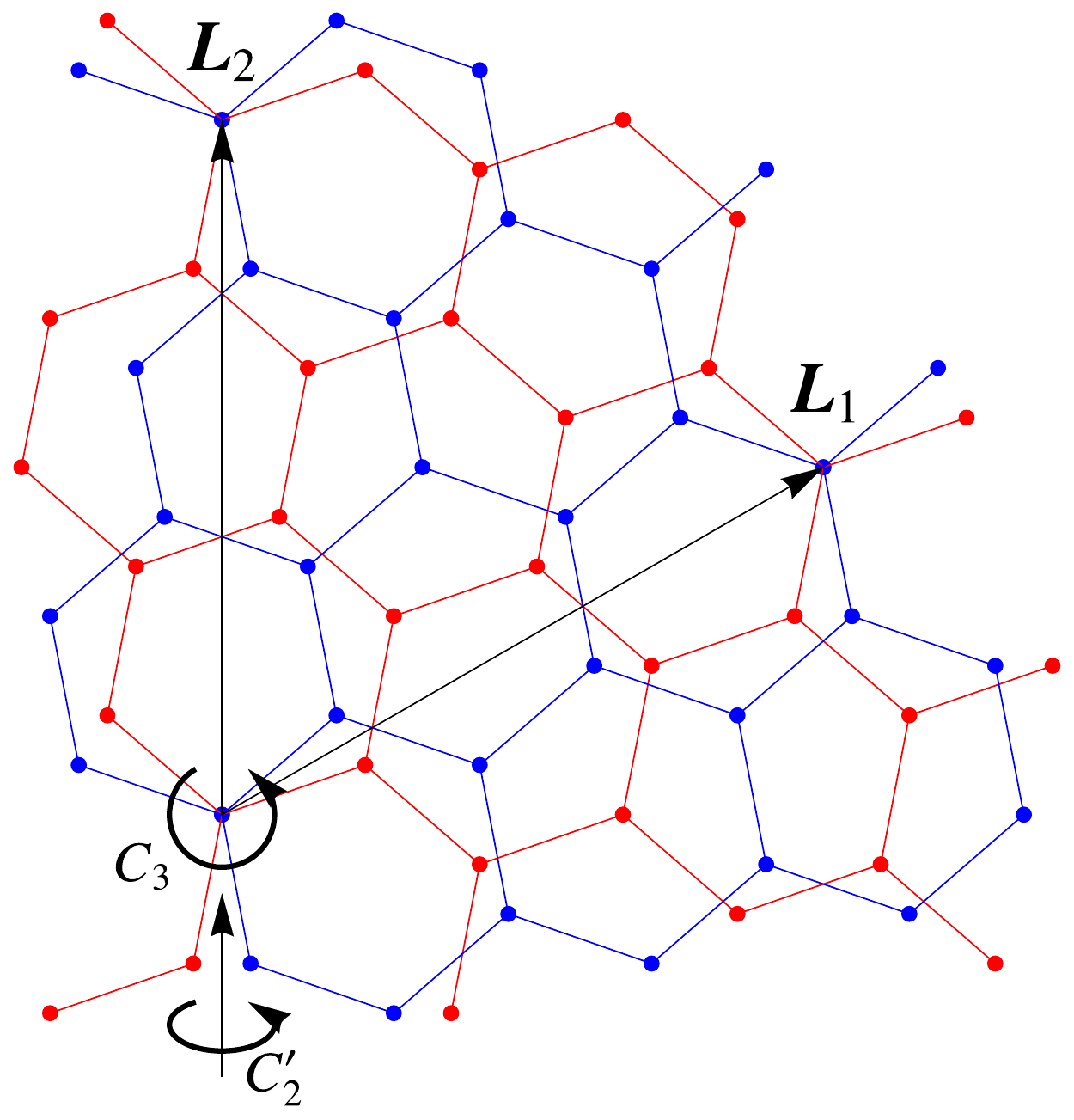}}
\subfigure[\label{Fig:lattice:Wyckoff}]{\includegraphics[width=0.48\columnwidth]{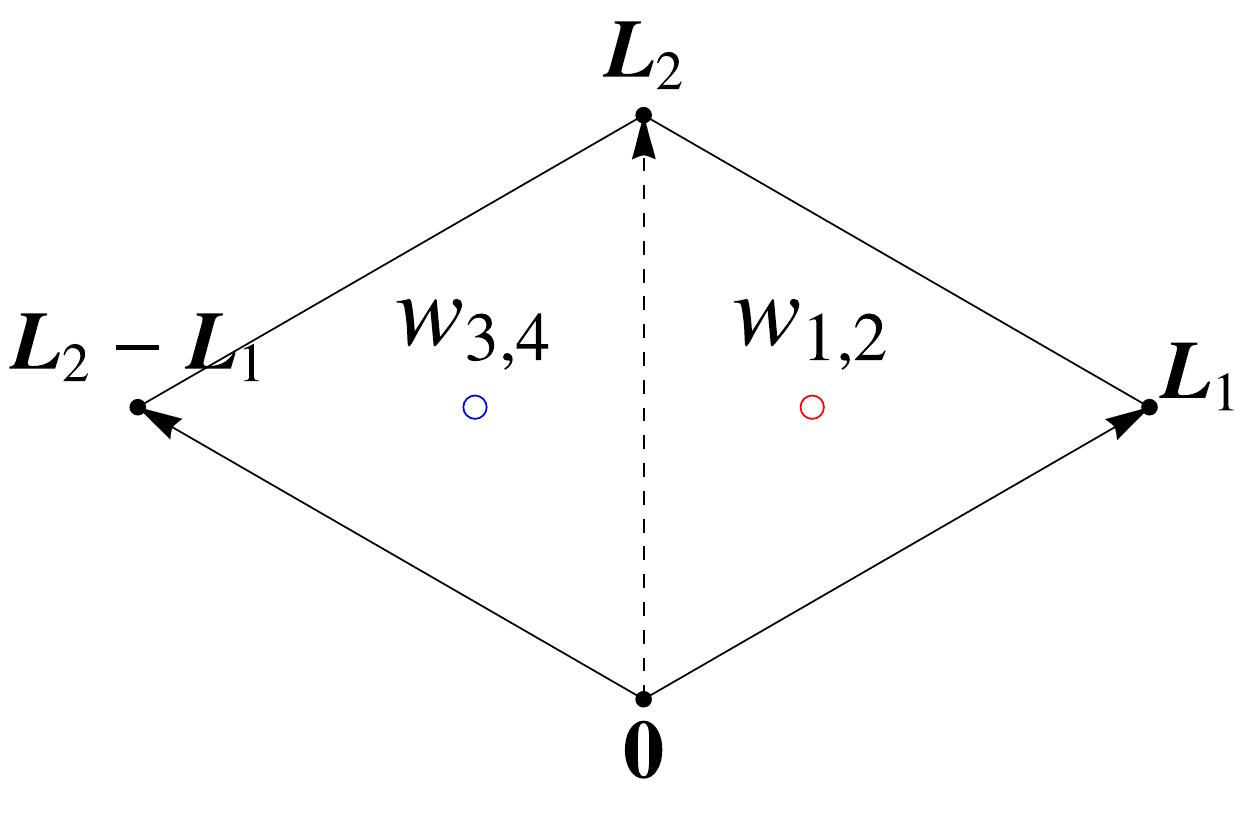}} 
\caption{ (Left) The superlattice of the twisted bilayer graphene. Blue (red) sites are the carbon atoms on the bottom (top) layers. The triangular lattice forms when the twisted angle is commensurate. The plot shows the lattice when $m = 2$, and $n = 1$.
(Right) The center of the local Wannier states. Black dots are the sites of the triangular superlattice. Red and blue dots are two nonequivalent Wyckoff sites, where the local Wannier states centered. In our construction, $w_1$ and $w_2$ are placed at one Wyckoff position, and $w_3$ and $w_4$ are placed at another position. Note that the Wychoff sites form an emergent honeycomb lattice.}
\label{Fig:lattice}
\end{figure}

Several theories have been proposed to address the insulating and superconducting phases~\cite{MacDonald, Balents, Liang, Senthil, Juricic, Scalettar, Baskaran, Phillips, Kivelson, Ma, Zhang, Das, Yang, Patrick, Heikkila, Spalek}. The closest to ours are Refs.~\cite{Senthil,Liang}. However, there are also important differences. In the theory of Ref.~\cite{Senthil}, the valley $U(1)$ symmetry -- and its spontaneous breaking -- plays an important role. Such valley symmetry, together with the product of $C_2$ and time reversal, is claimed to be an obstruction to building a tight binding model for the four narrow bands~\cite{Senthil}. In our microscopic construction we only have the three-fold rotation about the axis formed by the AA stacked carbon atoms ($C_3$), the two-fold rotation about the axis perpendicular to the two atoms ($C'_2$), and the time reversal symmetry (see Fig.~\ref{Fig:lattice:Superlattice}).
 We find the same group representations of the Bloch states at the high symmetry MBz points as conjectured in Ref.~\cite{Liang}. Although the WSs were not constructed explicitly in Ref.~\cite{Liang}, the WS symmetry was insightfully deduced and is in agreement with our findings. The three-peak structure of the WSs which we find explicitly (see Fig.~\ref{Fig:Wannier:1}) was also recognized in Ref.~\cite{Senthil} and dubbed `fidget spinner'.

\emph{Superlattice and band structure of the tBG}: For a commensurate twist angle the Morie pattern can be specified by two integers $(m, n)$, see e. g. ~\cite{Mele}. The primitive translation vectors $\fvec L_1 = m \fvec a_1 + n \fvec a_2$ and $\fvec L_2 = -n \fvec a_1 + (m + n)\fvec a_2$, where $a_1$ and $a_2$ are the primitive vectors of the single layer graphene lattice. As shown in Fig.~\ref{Fig:lattice}, the triangular super-lattice sites are the positions of $AA$ stacking. The point group symmetry operations form the $D_3$ group generated by $C_3$ and $C'_2$. This  leads to  nontrivial symmetry representations of the Bloch states at the high symmetry points in MBz, especially at $\fvec \Gamma$ ($\fvec k = 0$) and $\fvec K$ ($\fvec k = \frac{4\pi}{3 \fvec L_1^2} \fvec L_1$).

We calculate the band structure based the microscopic model of Ref.~\cite{Koshino}, which gives the values of the intralayer and the interlayer carbon-carbon tunneling amplitudes. Their tight binding Hamiltonian is written as
\begin{equation}
H  = \sum_{\fvec r_i , \fvec r_j} t(\fvec r_i - \fvec r_j) c_{\fvec r_i}^{\dagger} c_{\fvec r_j} \ , \label{Eqn:TBM_DFT}
\end{equation}
where $c_{\fvec r_i}$ and $c_{\fvec r_i}^{\dagger}$ are the annihilation and creation operators of the electron at the carbon site $\fvec r_i$. The detailed parameters are reproduced from Ref.~\cite{Koshino} in the supplementary material (SM) for completeness. The MBz contains three high symmetry points $\fvec \Gamma$, $\fvec K$ and $\fvec K'$. The time reversal symmetry (TRS) transform $\fvec K$ and $\fvec K'$ into each other and leaves $\fvec \Gamma$ invariant.

\begin{figure}[htbp]
\centering
\includegraphics[scale=0.5]{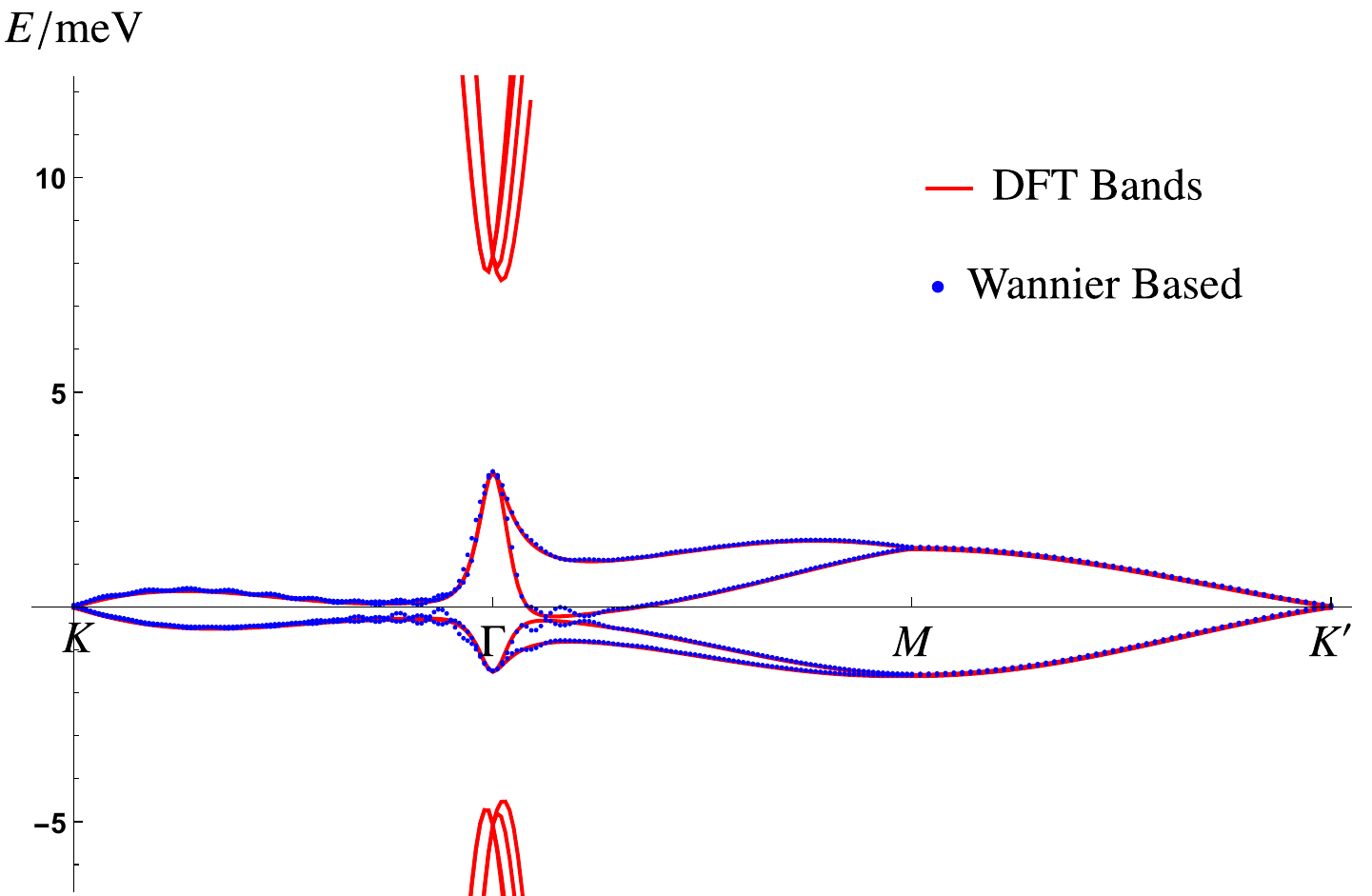} 
\caption{Red dots: the four narrow bands produced from the tight binding model with hopping parameters given in Ref.~\cite{Koshino}. Blue dots: the interpolated band structure generated by Wannier90.}
\label{Fig:DFTEne}
\end{figure}

\begin{widetext}
\begin{figure*}[htbp]
\centering
\subfigure[\label{Fig:Wannier:Bloch1}]{\includegraphics[width=0.5\columnwidth]{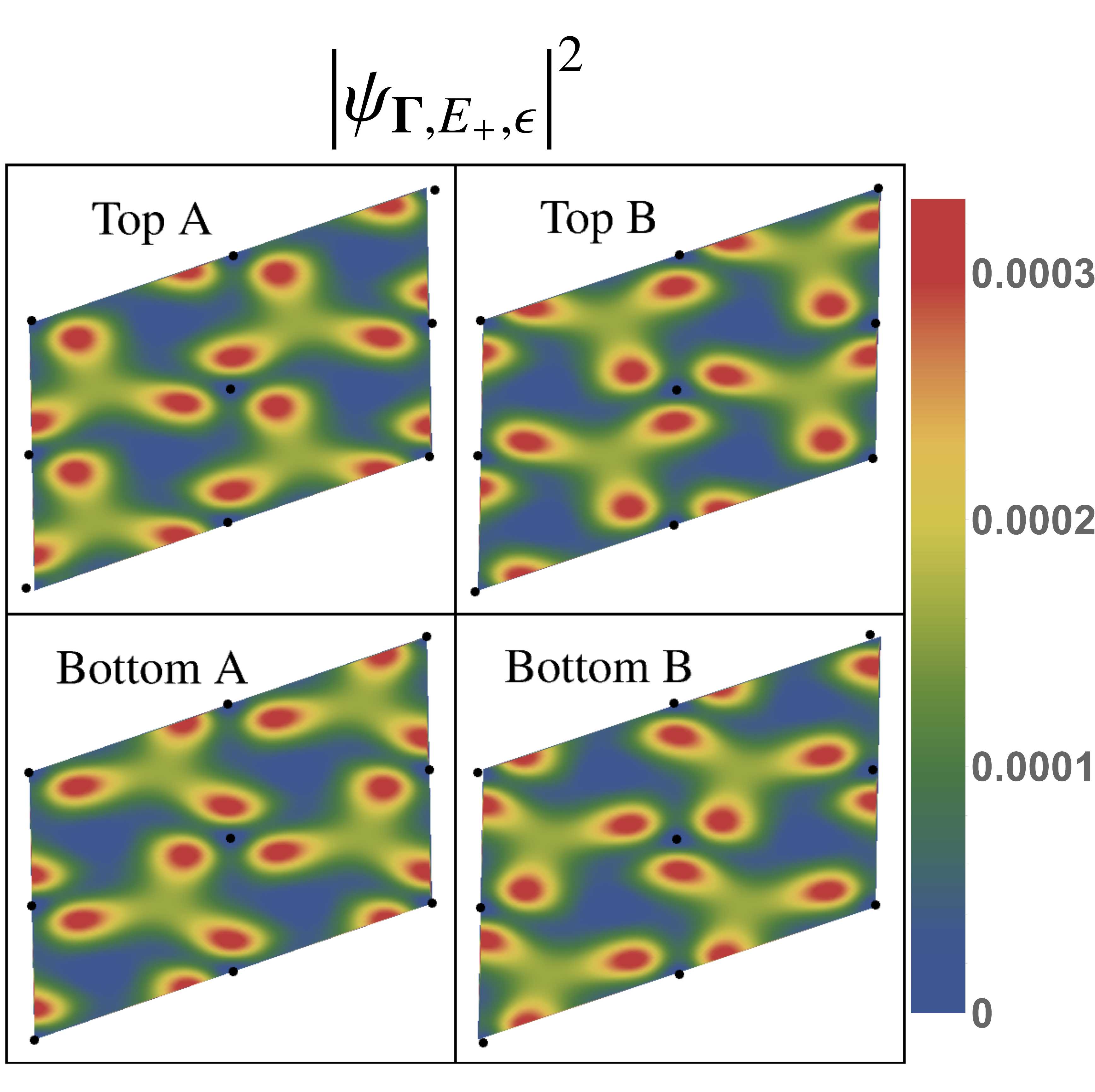}}
\subfigure[\label{Fig:Wannier:Bloch3}]{\includegraphics[width=0.5\columnwidth]{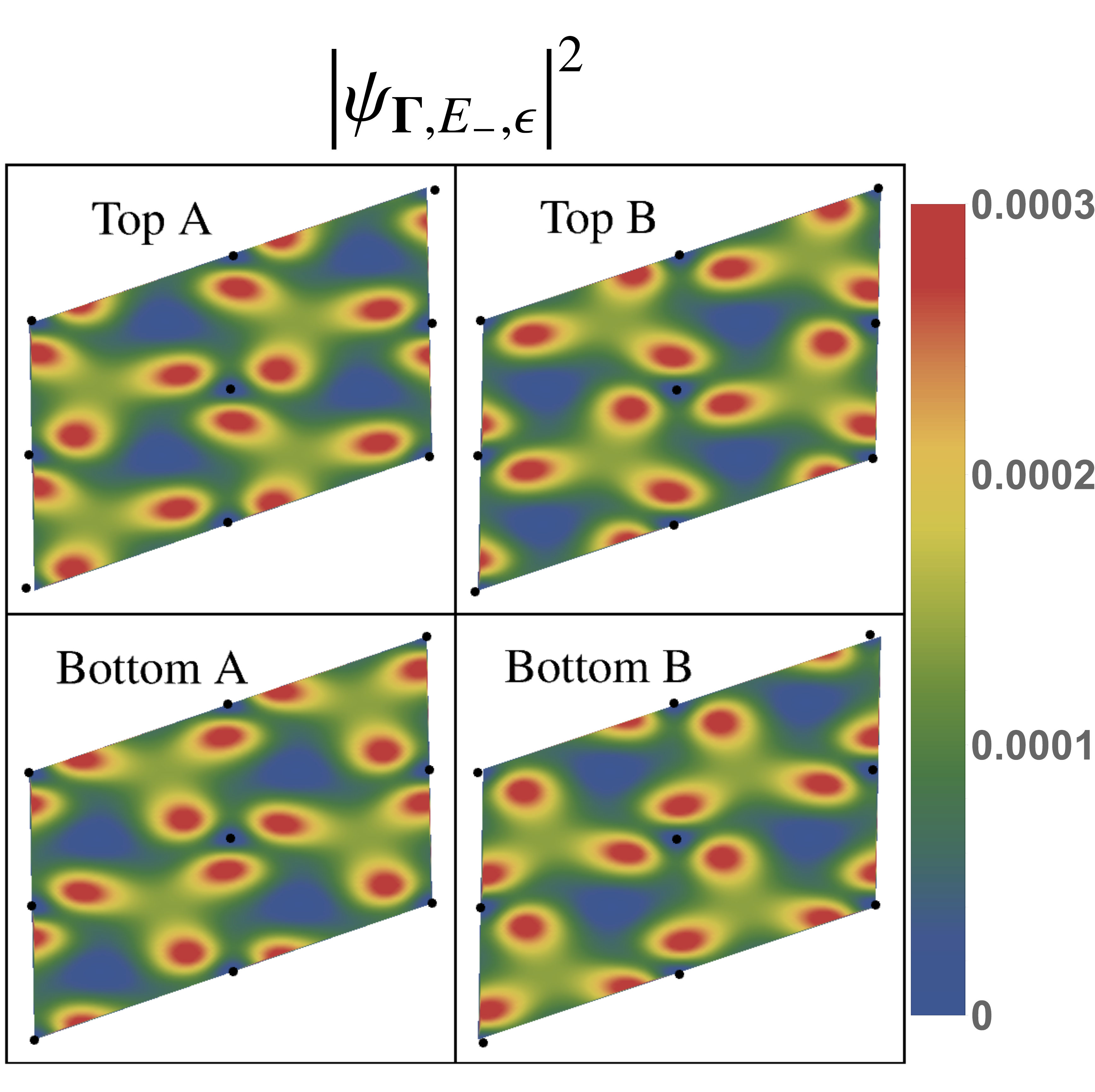}}
\subfigure[\label{Fig:Wannier:1}]{\includegraphics[width=0.85\columnwidth]{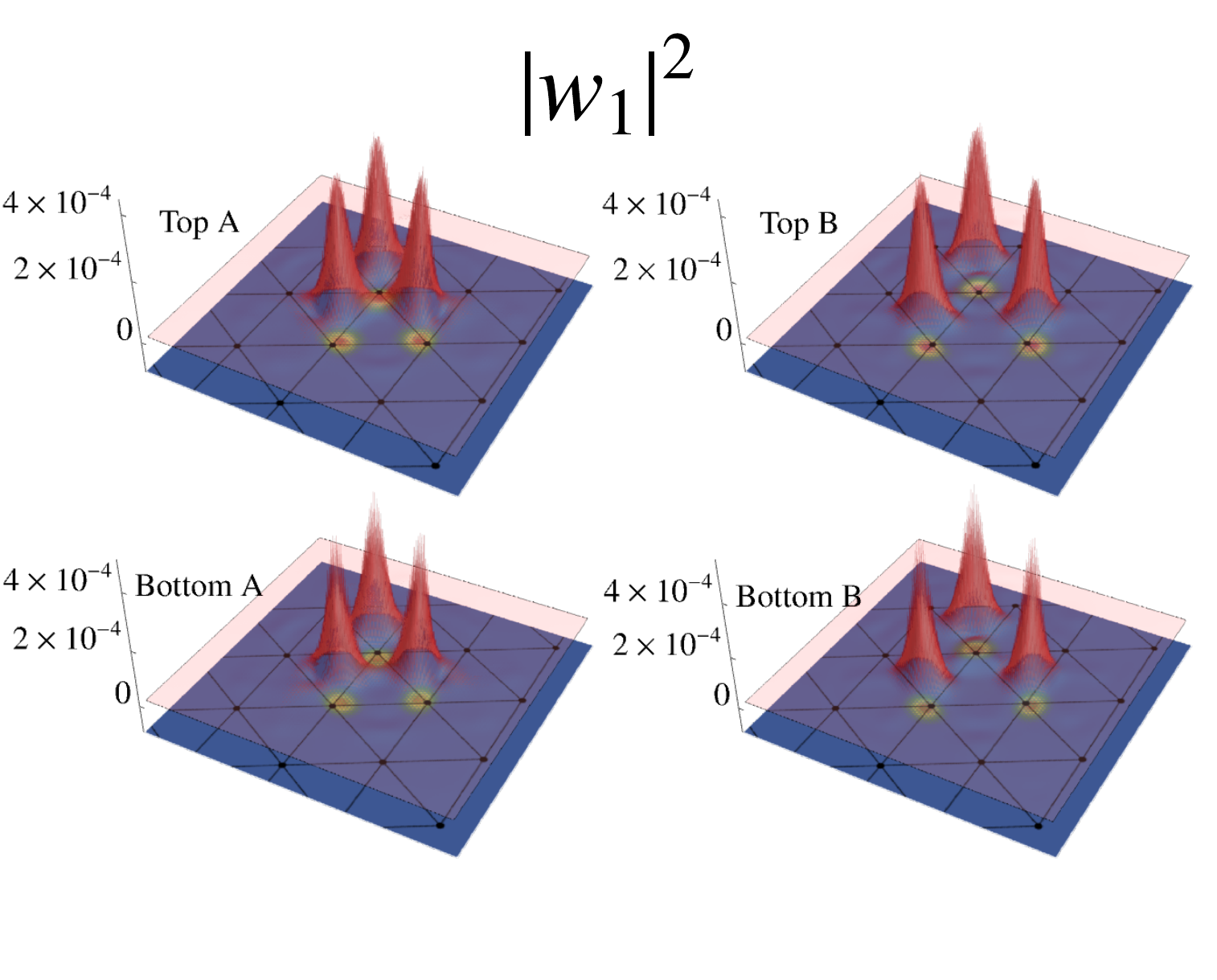}}
\caption{(a) \& (b) the square of the magnitude of the Bloch states $|\psi_{\fvec \Gamma, E_+,\epsilon}|^2$ and $|\psi_{\fvec \Gamma, E_-, \epsilon}|^2$, and (c) the Wannier state $|w_1|^2$ at (upper left) the top layer sublattice A, (upper right) the top layer sublattice B, (lower left) the bottom layer sublattice A, and (lower right) the bottom layer sublattice B.}
\label{Fig:Wannier}
\end{figure*}
\end{widetext}

As illustrated in Fig.\ref{Fig:DFTEne}, this model contains four narrow bands with very small bandwidths near the charge neutrality point where the zero of energy has been defined. Depending on the value of the twist angle, these four bands may or may not be separated by an energy gap from the other bands in the spectrum. When $m - n = \pm 1 \mod 3$~\cite{Mele}, at the $\fvec K$ point, two bands form a Dirac cone and the remaining two bands are split by a tiny gap ($ < 0.01$meV). These four Bloch states at $\fvec K$ form a two-dimensional representation ($E$) and two one-dimensional representations ($A_1$ and $A_2$) of the group $D_3$~\cite{Liang}, consistent with the degeneracy described above.
The Bloch states at the center of the MBz, $\fvec \Gamma$, are doubly-degenerate; the energy difference between the two pairs defines the (narrow) band width. The doublets are the two-dimensional representations ($E$) of the group $D_3$~\cite{Liang}. Using $\epsilon$ to represent the phase factor $\epsilon = \exp(i 2\pi/3)$~\cite{LandauQM}, we choose the two components of each doublet to transform as the eigenstates of $C_3$ with the eigenvalues of either $\epsilon$ or $\epsilon^*$, and label the four Bloch states at $\fvec \Gamma$ as $\psi_{\fvec \Gamma, E_{\pm}, \epsilon^{\pm1}}$. Here $E_{\pm}$ refers to the doublet with higher (lower) energy and $\epsilon^{\pm 1}$ refers to the component of the doublet which has the eigenvalue of $\epsilon$ ($\epsilon^*$) under $C_3$. While the two components of each doublet are the eigenstates of $C_3$, they transforms into each other under $C'_2$ and the TRS. We wish to stress that there is no simple transformation which relates the two doublets at different energy i.e. $\psi_{ {\fvec \Gamma}, E_{\pm}}$. This can be seen in Fig.~\ref{Fig:Wannier:Bloch1} and \ref{Fig:Wannier:Bloch3} where  $|\psi_{\fvec \Gamma}|^2$ are plotted.

\emph{Wannier states}: Our next step is to construct the localized WSs by applying the projection method \cite{Vanderbilt}. For this purpose, it is necessary that the four bands are separated by a gap from all others. The experiments of Ref.~\cite{Pablo1,Pablo2} determined that the closest simple commensurate values are $m = 30$ and $n = 31$. However, the four bands produced by Eqn.~\ref{Eqn:TBM_DFT} are gapped only near the band maximum, not near the band minimum; this is also seen in Ref.~\cite{Pablo1} Fig.~1. Such connection with the bands below contradicts the experimental finding that the four bands of interest are separated from either side by insulating states~\cite{Pablo1}. Therefore, we construct the WSs for the case of $m = 25$ and $n = 26$ (with the twist angle $\theta = 1.30^{\circ}$); the four bands are then separated by a gap on both sides. We expect that the values of the hopping parameters of the low energy Hamiltonian at the `magic' angle to be almost the same, and, importantly, can be fine tuned to it by slight modification. We confirmed that the quadratic band touching at $\fvec K$, which can be taken to be the defining property of the `magic' angle, can be realized in such a way.

As mentioned, it is crucial to identify the positions of the WSs. One naive choice is to place centers of all four states on the triangular Moire superlattice sites. With this option, WSs transform as
\begin{equation}
g | w_{i, \fvec R} \rangle = \sum_j | w_{j, g \fvec R} \rangle U_{ji}(g) \label{Eqn:SymmetryWannier}
\end{equation}
where $i, j = 1, \cdots 4$ are the indices of the WSs, $\fvec R$ is the position of the triangular super-lattice site, and $g$ is the symmetry  operation. The Bloch state $\psi_{i, \fvec k}$ is the linear superposition of the WSs. Under the same symmetry operation $g$, we find
\begin{align}
  & g | \psi_{i, \fvec k} \rangle  = g \sum_{\fvec R} e^{ i \fvec k \cdot \fvec R} |w_{i, \fvec R} \rangle = \sum_{\fvec R} e^{ i \fvec k \cdot \fvec R} |w_{j, g \fvec R} \rangle U_{ji}(g) \nonumber \\
  & = \sum_{\fvec R} e^{i g\fvec k \cdot g \fvec R} |w_{j, g \fvec R} \rangle U_{ji}(g) = | \psi_{j, g \fvec k} \rangle U_{ji}(g) \ . \label{Eqn:SymWannier1}
\end{align}
It is interesting to study the special case when the momentum is symmetry invariant, i.e.~$\fvec \Gamma$ and $\fvec K$ in the MBz. We immediately conclude that the Bloch states should transform as $U(g)$, and therefore, the Bloch states should transform in the \emph{same} way at $\fvec \Gamma$ and $\fvec K$. As we pointed out, the four Bloch states transform as two doublets at $\fvec \Gamma$, and \emph{one} doublet and \emph{two} singlets at $\fvec K$. This proves that the symmetry of the Bloch states cannot be reproduced if all the WSs are placed at the sites of the triangular super lattice.

The argument above suggests that the centers of the four WSs should be placed at non-equivalent sites (Wyckoff positions) to reproduce the symmetry representations at $\fvec \Gamma$ and $\fvec K$. A better choice is to place them at the centers of the equilateral triangles (Fig.~\ref{Fig:lattice:Wyckoff}), which form the dual honeycomb lattice~\cite{Liang,Senthil}. Note that each triangular superlattice unit cell contains two honeycomb lattice sites. The two WSs, $w_1$ and $w_2$, should be placed at one site, and  $w_3$ and $w_4$ at another site.

To illustrate the symmetry of the Wannier states, we start by modifying the Eq.~\eqref{Eqn:SymmetryWannier} for the dual honeycomb lattice~\cite{Sakuma},
\begin{equation}
  g | w_{i, \fvec R} \rangle = \sum_j | w_{j, g \fvec R + \fvec R'(g, i)} \rangle U_{ji}(g) \ , \label{Eqn:HoneycombWannier}
\end{equation}
where $\fvec R$ and $\fvec R'$ are still the \emph{triangular} lattice translation vectors, the latter depends only on $g$ and the WS index $i$. The Eq.~\ref{Eqn:SymWannier1} now takes the form~\cite{Sakuma}
\begin{equation}
  g | \psi_{i, \fvec k} \rangle = | \psi_{j, g \fvec k} \rangle e^{-i g \fvec k \cdot \fvec R'} U_{ji}(g) \ .
\end{equation}
Note that the extra phase factor, $e^{-i g\fvec k \cdot \fvec R'}$, now differentiates between $\fvec \Gamma$ (where it is $1$) and $\fvec K$ (in general nontrivial).
For $g=C_3$ and ${\bf k}=\fvec \Gamma$ the matrix $U$ must be diagonal i.e. all four WSs must be eigenstates of the $C_3$ followed by a lattice translation, with the same eigenvalues as those of $| \psi_{i, \fvec \Gamma} \rangle$. We therefore choose the $w_{1,4}$ and $w_{2,3}$ to have the eigenvalues $\epsilon$ and $\epsilon^*$, respectively.
Next, because $C_2'$ interchanges the two non-equivalent Wyckoff positions and the $C_3$ eigenvalues, we can set $C_2' w_1 = w_3$ and $C_2' w_2 = w_4$, see Fig.~\ref{Fig:lattice:Wyckoff}.
Finally, the time reversal symmetry does not change the position of the WSs, but it does conjugate the eigenvalue of $C_3$. Therefore, $\mathcal{T} w_1 = w_2$ and $\mathcal{T} w_3 = w_4$. These transformation rules together with translation symmetry enforce the symmetry of any low energy model.

As shown in Figs.~\ref{Fig:Wannier:Bloch1}-\ref{Fig:Wannier:Bloch3}, we found that the magnitudes of the Bloch states at $\fvec \Gamma$ display a smooth structure in real space when separated out by the layer and the microscopic carbon sublattice. This observation, along with the above considerations, suggests that a good initial ansatz for $w_1$ can be constructed as follows: first imagine placing a Gaussian-like cutoff centered at the first dual honeycomb site on $\psi_{\Gamma,E_+,\epsilon}$, but only on the top layer and sublattice A, and the bottom layer and sublattice B. The amplitudes at the top layer and the sublattice B, and the bottom layer and the sublattice A, are taken from the similarly cut-off $\psi_{\Gamma,E_-,\epsilon}$. This guarantees good overlap with the Bloch states. $C'_2$ now generates $w_3$, and the TRS generates $w_2$; when the TRS is applied to $w_3$ it finally gives $w_4$.

In the next step, we use the initial ansatz as an input to the Wannier90 program~\cite{Wannier90} with site symmetry enforced, on the $30\times 30$ $\fvec k$-space mesh, and after $200$ iterations obtain the four maximally localized WSs. Fig.~\ref{Fig:Wannier:1} shows the shape of the resulting $|w_1|^2$ on different layers and different sublattices. As seen, $w_1$ is well localized and centered around the dual honeycomb lattice site; it also displays three different peaks, located around the triangular lattice sites. This is consistent with the local density of states obtained by DFT calculations which are also peaked around the triangular lattice sites. We checked that the remaining WSs obtained in this way are related by the mentioned symmetry: $w_2=w^*_1$, $w_3 = C_2' w_1$ and $w_4 = w_3^*$.

\emph{Tight binding model}: The tight binding model based on the maximally localized WSs can be readily constructed. The on-site term must be of the form
\begin{equation}
  H_{onsite} = -\mu\sum_{\fvec R} \sum_{j = 1}^4 f_{j, \fvec R}^{\dagger} f_{j,\fvec R}
\end{equation}
where $f_{i,\fvec R}^{\dagger}$ and $f_{i,\fvec R}$ are the creation and annihilation operators of the WSs $w_{i, \fvec R}$. This is because $C_3$ prohibits mixing between $w_1$ and $w_2$, and $w_3$ and $w_4$; $C_2'$ and TRS then force a single real parameter $\mu$. In contrast, there are two such parameters in Ref.~\cite{Senthil}. The nearest neighbor hopping term is between $w_{1,2;\fvec R_i}$ and the neighboring $w_{3,4; \fvec R_j}$. The most general term allowed by symmetry is
\begin{align}
  H_n & = \sum_{\fvec R} \left\{ t_1  f_{1,\fvec R}^{\dagger} \left(  f_{4,\fvec R}  + f_{4, \fvec R + \fvec L_1} + f_{4, \fvec R + \fvec L_1 - \fvec L_2} \right) + \right. \nonumber \\
  &  t_1  f_{2\,\fvec R}^{\dagger} \left( f_{3,\fvec R}  + f_{3, \fvec R + \fvec L_1} + f_{3, \fvec R + \fvec L_1 - \fvec L_2} \right) \nonumber \\
  & t_1' f_{1\,\fvec R}^{\dagger} \left( f_{3,\fvec R}  + \epsilon f_{3, \fvec R + \fvec L_1} + \epsilon^* f_{3, \fvec R + \fvec L_1 - \fvec L_2} \right) + \nonumber \\
  & \left. t_1'^* f_{1\,\fvec R}^{\dagger} \left( f_{3,\fvec R}  + \epsilon^* f_{3, \fvec R + \fvec L_1} + \epsilon f_{3, \fvec R + \fvec L_1 - \fvec L_2} \right) \right\} + h.c.
\end{align}
where $t_1$ is real and $t_1'$ are, in general, a complex number (see SM). The structure of WSs, seen in Fig.~\ref{Fig:Wannier:1}, suggests that the overlap of the next and even the next-next nearest neighboring WSs is sizable, and thus cannot be neglected even in the minimal model. The detailed analysis of the symmetry constraints on the further range hopping is now straightforward. It is presented in the SM, where we also study the quality of the fit as a function of the hopping range.  The blue points in Fig.~\ref{Fig:DFTEne} shows the interpolated band structure obtained from the Wannier90 program using all the hoppings.  The agreement with the the microscopic tight binding model based on DFT calculations (red solid lines) suggests that our low energy model accurately reproduces the main physics.

\emph{Conclusion}: In this paper, we presented a method for constructing symmetry adapted  maximally localized Wannier functions and the corresponding low energy model for the four narrow bands of the tBG near the `magic' angle. The WSs have three peaks around the Moire triangular lattice sites, but are centered at the dual honeycomb lattice sites. They form non-trivial representations of the site symmetry group. Our model provides a firm basis for further study of the many-body effects.

\begin{acknowledgments}
  We thank Yuan Cao, Valla Fatemi, Liang Fu, Pablo Jarillo-Herrero, Leonid Levitov, Senthil Todardi, Ashvin Vishwanath and Y.~Y.~Zhao for fruitful discussions. JK was supported by the National High Magnetic Field Laboratory through NSF Grant No.~DMR-1157490 and the State of Florida.  O. V. was supported by NSF DMR-1506756.
\end{acknowledgments}

\vspace{0.5cm}

\widetext

\begin{center}
\textbf{\large{}{}{}{}{}{}Supplementary material for ``Symmetry, maximally localized Wannier states, and low energy model for the twisted bilayer graphene narrow bands''}{\large{}{}{}{}{}{}
}
\par\end{center}
\setcounter{equation}{0} \setcounter{figure}{0} \setcounter{table}{0}

\global\long\def\theequation{S\arabic{equation}}

\global\long\def\thetable{S\arabic{table}}
 \global\long\def\thefigure{S\arabic{figure}}


\section{Microscopic Tight Binding Model}
We use the same model as in Ref.~\cite{Koshino1} to produce the band structure. The Hamiltonian is
\begin{align}
H  & = \sum_{\fvec R_i , \fvec R_j} t(\fvec R_i - \fvec R_j) c_{\fvec R_i}^{\dagger} c_{\fvec R_j} \ , & \mbox{with} \quad & t(\fvec d)  = -V_{pp\pi} \left[ 1 - \left( \frac{\fvec d \cdot \fvec e_z}{d} \right)^2 \right] - V_{pp\sigma} \left( \frac{\fvec d \cdot \fvec e_z}{d} \right)^2 \nonumber \\
V_{pp\pi} & = V_{pp\pi}^0 \exp\left( - \frac{d - a_0}{\delta} \right) & & V_{pp\sigma}  = V_{pp\sigma}^0 \exp\left( - \frac{d - a_0}{\delta} \right)
\end{align}
where we the first and the second term in the Hamiltonian $H$ are for the intralayer and the interlayer tunneling, respectively. We set $V_{pp\pi}^0 = -2.7$eV, $V_{pp\sigma}^0 = 0.48$eV.
$a_0 = 0.142$nm is the distance between the two nearest neighbor carbon atoms on the same layer.  The decay length for the hopping is $\delta = 0.319 a_0$. The hopping with $d > 4a_0$ is exponentially small and thus is neglected in the model. The band structure produced by this model is illustrated as red solid lines in Fig.~\ref{Fig:DFTEne}.

\section{Projection Method}
In this section, we will explain the projection method we used to produce the localized WSs as the input of the Wannier90 program. We follow the approach in Ref.~\cite{Vanderbilt1}. As explained in the text, we first construct the trial functions $| f_i \rangle$ ($i = 1, \cdots 4$), which transform in the same way as the WSs. These trial states are not necessarily orthogonal or normalized. For the Bloch states $|\psi_{i, \fvec k} \rangle$, we define the matrix $A(\fvec k)_{ij} = \langle \psi_{i, \fvec k} | f_j \rangle$. The states
\[  | \phi_{i,\fvec k} \rangle = \sum_j |\psi_{j, \fvec k} \rangle \langle \psi_{j, \fvec k} | f_i \rangle  = \sum_j |\psi_{j, \fvec k} \rangle A_{ji}(\fvec k) \]
are smooth in $\fvec k$, because the arbitrary $\fvec k$-dependent phase cancels in the projector. Smoothness in $\fvec k$ is required in order for WSs to be localized in real space. However, they are not orthonormal. To construct the orthonormalized $\fvec k$-smooth Bloch-like states, we define the matrix $S(\fvec k) = A^{\dagger}(\fvec k) A(\fvec k)$, and
\[ | \tilde{\psi}_{i, \fvec k} \rangle = \sum_j | \phi_{j, \fvec k} \rangle S^{-1/2}_{ji}(\fvec k) =   \sum_j | \psi_{j, \fvec k} \rangle  \left( A(\fvec k) S^{-1/2}(\fvec k) \right)_{ji} \ . \]
In practice, we apply the singular value decomposition to the matrix $A(\fvec k) = U(\fvec k) D(\fvec k) V^{\dagger}(\fvec k)$, where the matrices $U(\fvec k)$ and $V(\fvec k)$ are unitary, and $D(\fvec k)$ is diagonal. It is easy to show that $A(\fvec k) S^{-1/2}(\fvec k) = U(\fvec k) V^{\dagger}(\fvec k)$ is unitary, and thus, $| \tilde{\psi} \rangle$ is orthogonal and normalized. With the projection method, the WSs are
\[ | w_{i,\fvec R} \rangle = \int\frac{\rmd^d \fvec k}{(2\pi)^d} | \tilde{\psi}_{i, \fvec k} \rangle e^{-i \fvec k \cdot \fvec R} \ . \]
We use the above as an input to Wannier90 program with site symmetry enforced to obtain the maximally localized symmetry adapted WSs.

\section{Low energy Tight Binding model}
\subsection{Symmetry Constraints}
In this subsection, we discuss the most general form of the hopping amplitudes allowed by symmetry. As mentioned in the main text, the symmetries are $C_3$, $C_2'$, and the TRS. Since we have already explained the constraints on the on-site hybridization term in the main text, we first study the next term which is the hopping between $w_{1,2;\fvec R_i}$ and the neighboring $w_{3,4; \fvec R_j}$. Thus, the Hamiltonian is of the form
\begin{align}
  H_n = \sum_{\fvec R} \sum_{\substack{i = 1,2\\ j = 3, 4}}  f_{i,\fvec R}^{\dagger} \left( t^n_{ i j} f_{j, \fvec R}  + t^{n'}_{ij} f_{j, \fvec R + \fvec L_1 - \fvec L_2}  + t^{n''}_{i j} f_{j, \fvec R + \fvec L_1} \right) + h.c.
\end{align}
It should be invariant under all the symmetry transformations. First, consider $C_3$ which brings WSs $w_{\fvec R}$ into $w_{\fvec R'}$ in a different unit cell. In addition, $w_1$ and $w_4$ have the eigenvalue of $\epsilon$, and $w_2$ and $w_3$ have the eigenvalue of $\epsilon^*$. The $C_3$ invariance of the Hamiltonian forces
\[ t^n_{14} = t^{n'}_{14} = t^{n''}_{14} \ , \qquad t^n_{23} = t^{n'}_{23} = t^{n''}_{23} \ , \qquad t_{13}^n = \epsilon t_{13}^{n'} = \epsilon^* t_{13}^{n''} \ , \quad \mbox{and} \quad  t_{24}^n = \epsilon^* t_{24}^{n'} = \epsilon t_{24}^{n''} \ .  \]
$C_2'$ transforms $w_1 \leftrightarrow w_3$ and $w_2 \leftrightarrow w_4$, and brings $\fvec L_1 \rightarrow \fvec L_2 - \fvec L_1$, and $\fvec L_2 \rightarrow \fvec L_2$. Combined with the hermiticity of the Hamiltonian, the $C_2'$ invariance leads to
$t_{14}^n = \left( t_{23}^n \right)^*$, $t_{13}^n = \left( t_{13}^n \right)^*$ and  $t_{24}^n = \left( t_{24}^n \right)^*$.
Finally, the TRS enforces $t_{13}^n = \left( t_{24}^n \right)^*$.

Combining all constraints, we set $t_{13}^n = t_1$ and $t_{14}^n = t_1'$, where $t_1$ is real and $t_1'$ is in general, a complex number. Thus, the nearest neighbor hopping term can be written as
\begin{align}
  H_n = \sum_{\fvec R} \begin{pmatrix} f_{1,\fvec R} \\ f_{2,\fvec R} \end{pmatrix}^{\dagger} \left\{ \begin{pmatrix} t_1 & t_1' \\ \left( t_1' \right)^* & t_1  \end{pmatrix} \begin{pmatrix} f_{3,\fvec R} \\ f_{4,\fvec R} \end{pmatrix} +
  \begin{pmatrix} \epsilon^* t_1 & t_1' \\ \left( t_1' \right)^* & \epsilon t_1  \end{pmatrix} \begin{pmatrix} f_{3,\fvec R + \fvec L_1 - \fvec L_2} \\ f_{4,\fvec R + \fvec L_1- \fvec L_2} \end{pmatrix} +
  \begin{pmatrix} \epsilon t_1 & t_1' \\ \left( t_1' \right)^* & \epsilon^* t_1  \end{pmatrix} \begin{pmatrix} f_{3,\fvec R + \fvec L_1 } \\ f_{4,\fvec R + \fvec L_1} \end{pmatrix}  \right\} + h.c.
\end{align}
It seems that $t_1'$ is a complex number. If we apply a gauge transformation $w_{1,3} \rightarrow e^{i\theta} w_{1,3}$ and $w_{2,4} \rightarrow e^{-i\theta} w_{2,4}$, the hopping constant $t_1$ is invariant but $t_1' \rightarrow e^{2i\theta} t_1'$. Thus, the phase of $t_1'$ can be always removed by choosing a particular gauge of the WSs. Therefore, there are only two free parameters for the nearest neighbor hopping~\cite{Liang1}.

Next, consider the next nearest neighbor hopping $H_{nn}$:
\begin{align}
  H_{nn} & =  \sum_{\fvec R} \left( \sum_{i,j = 1}^2 +  \sum_{i,j = 3}^4 \right) f_{i,\fvec R}^{\dagger} \left( t^{nn,1}_{i j} f_{j, \fvec R + \fvec L_1} + t^{nn,2}_{i j} f_{j, \fvec R + \fvec L_2 - \fvec L_1} + t^{nn,3}_{ij} f_{j, \fvec R - \fvec L_2}  \right) + h.c. \label{EqnS:NNHopping}
\end{align}
Let us first consider the symmetry constraints on $t_{ij}^{nn}$ when $i, j = 1, 2$. The $C_3$ invariance enforces
\[ t_{11}^{nn,1} = t_{11}^{nn,2} = t_{11}^{nn,3} \ , \quad  t_{22}^{nn,1} = t_{22}^{nn,2} = t_{2}^{nn,3} \ , \quad t_{12}^{nn,1} = \epsilon t_{12}^{nn,2} = \epsilon^* t_{12}^{nn,3} \ , \quad t_{21}^{nn,1} = \epsilon^* t_{21}^{nn,2} = \epsilon t_{21}^{nn,3} \ . \]
The TRS leads to
\[ t_{22}^{nn,1} = \left( t_{11}^{nn,1} \right)^*  \ , \quad t_{21}^{nn,1} = \left( t_{12}^{nn,1} \right)^* \ . \]
The hopping constants $t_{ij}^{nn}$ ($i, j = 3$ or $4$) can be obtained by applying $C_2'$ symmetry operation. Therefore, the next-nearest neighbor hopping can be described by two complex numbers $t_2 = t_{11}^{nn, 1}$ and $t_2' = t_{12}^{nn,1}$. The general form is
\begin{align}
  H_{nn} = & \sum_{\fvec R} \begin{pmatrix} f_{1,\fvec R} \\ f_{2,\fvec R} \end{pmatrix}^{\dagger} \left\{ \begin{pmatrix}
     t_2 & t_2' \\ (t_2')^* & t_2^*
   \end{pmatrix} \begin{pmatrix} f_{1,\fvec R + \fvec L_1} \\ f_{2,\fvec R + \fvec L_1} \end{pmatrix} +  \begin{pmatrix}
      t_2 & \epsilon^* t_2' \\ \epsilon (t_2')^* &  t_2^*
   \end{pmatrix}\begin{pmatrix} f_{1,\fvec R + \fvec L_2 - \fvec L_1} \\ f_{2,\fvec R + \fvec L_2 - \fvec L_1} \end{pmatrix} +  \begin{pmatrix}
      t_2 & \epsilon t_2' \\ \epsilon^* (t_2')^* &  t_2^*
   \end{pmatrix} \begin{pmatrix} f_{1,\fvec R - \fvec L_2 } \\ f_{2,\fvec R - \fvec L_2} \end{pmatrix} \right\} + \nonumber \\
   & \begin{pmatrix} f_{3,\fvec R} \\ f_{4,\fvec R} \end{pmatrix}^{\dagger} \left\{ \begin{pmatrix}
     t_2 & t_2' \\ (t_2')^* & t_2^*
   \end{pmatrix}\begin{pmatrix} f_{3,\fvec R + \fvec L_2 - \fvec L_1} \\ f_{4,\fvec R + \fvec L_2 - \fvec L_1} \end{pmatrix} +  \begin{pmatrix}
      t_2 & \epsilon^* t_2' \\ \epsilon (t_2')^* &  t_2^*
   \end{pmatrix}\begin{pmatrix} f_{3,\fvec R + \fvec L_1} \\ f_{4,\fvec R + \fvec L_1} \end{pmatrix} +  \begin{pmatrix}
      t_2 & \epsilon t_2' \\ \epsilon^* (t_2')^* &  t_2^*
   \end{pmatrix}\begin{pmatrix} f_{3,\fvec R - \fvec L_2 } \\ f_{4,\fvec R - \fvec L_2} \end{pmatrix} \right\}
\end{align}

Finally, we consider the symmetry constraints on the next-next nearest neighbor hopping, with the most general form of
\begin{align}
  H_{nnn} = \sum_{\fvec R} \sum_{\substack{i = 1,2\\ j = 3, 4}}  f_{i,\fvec R}^{\dagger} \left( t^{nnn,1}_{ i j} f_{j, \fvec R + 2\fvec L_1 - \fvec L_2} + t^{nnn,2}_{i j} f_{j, \fvec R + \fvec L_2}  + t^{nnn,3}_{ij} f_{j, \fvec R  - \fvec L_2} \right) + h.c.
\end{align}
The constraints are very similar to the one for the nearest neighbor hopping. We found $C_3$ enforces
\[ t_{14}^{nnn,1} =   t_{14}^{nnn,2} =  t_{14}^{nnn,3} \ , \quad  t_{23}^{nnn,1} =   t_{23}^{nnn,2} =  t_{23}^{nnn,3} \ , \quad t_{13}^{nnn,1} = \epsilon  t_{13}^{nnn,2} = \epsilon^* t_{13}^{nnn,3} \ , \quad t_{24}^{nnn,1} = \epsilon^*  t_{24}^{nnn,2} = \epsilon t_{24}^{nnn,3} \ , \]
Combined with the hermiticity of the Hamiltonian, $C_2'$  leads to
\[  t_{13}^{nnn,1} = \left( t_{13}^{nnn,1} \right)^* \ , \quad  t_{24}^{nnn,1} = \left( t_{24}^{nnn,1} \right)^* \ , \quad t_{14}^{nnn,1} = \left(  t_{23}^{nnn,1}   \right)^* \ . \]
The TRS puts an additional constraint $t_{13}^{nnn,1} = \left( t_{24}^{nnn,1} \right)^*$. Thus, we can introduce one real $t_3 = t_{13}^{nnn,1}$ and one complex $t_3' = t_{14}^{nnn,1}$ parameters for the next-next nearest neighbor hopping. The Hamiltonian takes the form
\begin{align}
  H_{nnn} & = \sum_{\fvec R} \begin{pmatrix} f_{1,\fvec R} \\ f_{2,\fvec R} \end{pmatrix}^{\dagger} \left\{ \begin{pmatrix} t_3 & t_3' \\ \left( t_3' \right)^* & t_3  \end{pmatrix} \begin{pmatrix} f_{3,\fvec R + 2\fvec L_1 - \fvec L_2} \\ f_{4,\fvec R + 2\fvec L_1 - \fvec L_2} \end{pmatrix} +
  \begin{pmatrix} \epsilon^* t_3 & t_3' \\ \left( t_3' \right)^* & \epsilon t_3  \end{pmatrix} \begin{pmatrix} f_{3,\fvec R + \fvec L_2} \\ f_{4,\fvec R + \fvec L_2} \end{pmatrix} +
  \begin{pmatrix} \epsilon t_3 & t_3' \\ \left( t_3' \right)^* & \epsilon^* t_3  \end{pmatrix} \begin{pmatrix} f_{3,\fvec R - \fvec L_2} \\ f_{4,\fvec R - \fvec L_2} \end{pmatrix}  \right\} + h.c.
\end{align}
The symmetry constraints on further range hopping can be worked out in the same way.

\subsection{Values of the Hopping Constants}
The most general tight binding Hamiltonian is of the form
\beq
H = \sum_{\fvec R, \fvec r} \sum_{i, j = 1}^{4} t_{ij, \fvec r} f_{i, \fvec R}^{\dagger} f_{j, \fvec R + \fvec r}
\eeq
where both $\fvec R$ and $\fvec r$ are the triangular lattice vectors. The hooping constants $t$ are indexed by two WS indices $i$ and $j$, and the lattice vector $\fvec r$. The numerical value of the hopping constant can be obtained from the energy of the Bloch states and the transformation between the WS and the Bloch states. Suppose that
\begin{align}
  | w_{i, \fvec R} \rangle & = \int \frac{\rmd^d \fvec k}{(2\pi)^d} |\psi_{j, \fvec k} \rangle  e^{-i \fvec k \cdot \fvec R} U_{ji}(\fvec k) \\
  t_{ij,\fvec r} & = \langle w_{i, \fvec R} | H |  w_{j, \fvec R + \fvec r} \rangle  = \sum_{i', j'} \int \frac{\rmd^d \fvec k \rmd^d \fvec k'}{(2\pi)^{2d}} e^{i \fvec k' \cdot \fvec R} U^*_{i' i}(\fvec k') \langle \psi_{i'}(\fvec k') | H | \psi_{j',\fvec k} \rangle e^{-i \fvec k \cdot (\fvec R + \fvec r)} U_{j' j}(\fvec k)  \nonumber \\
  & = \int \frac{\rmd^d \fvec k}{(2\pi)^d} U_{i' i}^*(\fvec k) \epsilon_{i'}(\fvec k) U_{i' j}(\fvec k) e^{-i \fvec k \cdot \fvec r}
\end{align}

In this subsection, we list the values of the hopping constants up to $|\fvec r| = 3 |\fvec L_1|$. For notation convenience, we write $\fvec r$ as two numbers $\fvec r = (a, b)$, meaning $\fvec r = a \fvec L_1 + b \fvec L_2$. Note that the , the TRS transforms $w_1 \rightarrow w_2$, and $w_3 \leftrightarrow w_4$. Thus, it enforces several constraints, e.g. $t_{12, \fvec r} = t_{21, \fvec r}^*$, $t_{13, \fvec r} = t_{24, \fvec r}^*$, etc. In the tables below, for notation simplicity, we only list part of the hopping constants; others can be obtained from the constraints due to hermiticity of the Hamiltonian and the TRS.

Here, we separate the hopping constants into two different tables. Tab.~\ref{TabS:InterHoppingWannier90} is for  the hoppings between $w_{1,2}$ and $w_{3,4}$, and Tab.~\ref{TabS:IntraHoppingWannier90} is for the hoppings among $w_{1,2}$ themselves, and $w_{3,4}$ themselves.

\begin{table}[h]
 \centering %
 \begin{tabular}{|c|c|}
  \hline
  $t_{13,\fvec 0} = \epsilon t_{13, (1, -1)} = \epsilon^* t_{13, (1, 0)}   $ &  $0.0831$ \\ \hline
  $t_{14, \fvec 0} = t_{14, (1, 0)} =  t_{14, (1, -1)}  $ & $0.0380 + 0.2603 i$ \\ \hline   \hline
  $t_{13, (2, -1)} = \epsilon t_{13, (0, 1)} =  \epsilon^* t_{13, (0, -1)} $   & -0.0853 \\ \hline
  $ t_{14,(2, -1)} = t_{14,(0, 1)} = t_{14,(0, -1)} $  &  $-0.0916 - 0.2868 i$  \\ \hline \hline
  $t_{13,(-1,0)} = t_{13, (-1, 1)}^* = \epsilon^* t_{13,(1,-2)}^*  = \epsilon^* t_{13,(1,1)}  = \epsilon t_{13,(2,-2)}  = \epsilon t_{13, (2,0)}^*  $  & $0.0299 -0.0279 i$ \\ \hline
  $ t_{14,(-1,0)} = t_{14,(-1, 1)} = t_{14, (1, -2)} = t_{14, (1,1)} = t_{14, (2,-2)} = t_{14, (2, 0)}$  &  $0.0339 + 0.0222 i$ \\ \hline \hline
  $t_{13, (-1, -1)} = \epsilon t_{13, (3, -2)} = \epsilon^* t_{13, (0,2)} = t_{13, (-1, 2)}^* = \epsilon^* t_{13, (0, -2)}^* = \epsilon t_{13, (3, -1)}^*$   &   $-0.0293 + 0.0009 i$ \\ \hline
  $ t_{14, (-1, -1)} = t_{14, (-1, 2)} = t_{14, (0, -2)} = t_{14, (0, 2)} = t_{14, (3, -2)} = t_{14, (3, -1)}$   &  $-0.0089 + 0.0112 i$  \\ \hline \hline
   $t_{13, (-2, 1)} = \epsilon t_{13, (2, -3)} = \epsilon^* t_{13, (2, 1)}$  & $0.0280$  \\ \hline
   $ t_{14, (-2, 1)} = t_{14, (2, -3)} = t_{14, (2, 1)} $ & $0.0021 - 0.0101 i$ \\ \hline \hline
   $t_{13, (-2, 0)} = \epsilon t_{13, (3, -3)} = \epsilon^*  t_{13, (1, 2)} = t_{13, (-2, 2)}^* = \epsilon^* t_{13, (1, -3)}^* = \epsilon t_{13, (3, 0)}^* $ & $ 0.0040 + 0.0256 i $ \\ \hline
   $ t_{14, (-2, 0)} = t_{14, (-2, 2)} = t_{14, (1, -3)} = t_{14, (1, 2)} = t_{14, (3, -3)} = t_{14, (3, 0)}$  &   $ 0.0131 + 0.0345 i $   \\ \hline   \hline
       $ t_{13, (4, -2)} = \epsilon t_{13, (-1, 3)} = \epsilon^*  t_{13, (-1, -2)}$   &  $ -0.0359 $  \\ \hline
    $ t_{14, (4, -2)} = t_{14, (-1, 3)} = t_{14, (-1, -2)} $  &   $ -0.0154 - 0.0398 i$ \\ \hline  \hline
    $ t_{13,(-2, -1)} = \epsilon t_{13, (4, -3)} = \epsilon^* t_{13, (0, 3)} = t_{13,(-2,3)}^* = \epsilon^* t_{13, (0, -3)} = \epsilon t_{13, (4, -1)}^* $    &   $-0.0107 - 0.0040 i$ \\ \hline
    $ t_{14,(-2, -1)} = t_{14, (4, -3)} = t_{14, (0, 3)} = t_{14, (-2, 3)} = t_{14, (0, -3)} = t_{14, (4, -1)} $    &  $ 0.0012 + 0.0099 i $ \\ \hline
\end{tabular}
 \caption{The hopping constants between $w_{1,2}$ and $w_{3,4}$. All the numbers are in the units of meV.}
\label{TabS:InterHoppingWannier90}
\end{table}

\begin{table}[h]
 \centering %
 \begin{tabular}{|c|c|}
  \hline
  $ t_{11, (1, 0)} =  t_{11, (-1, 1)} =  t_{11, (0, -1)} = t_{33, (1, 0)}  = t_{33, (-1, 1)} =  t_{33, (0, -1)}$ & $-0.0023 - 0.0161 i$ \\ \hline
  $t_{12, (1, 0)} = \epsilon t_{12, (-1, 1)} =  \epsilon^* t_{12, (0, -1)} = t_{34, (-1, 1)} =  \epsilon t_{34, (1, 0)} = \epsilon^* t_{34, (0, -1)}  $ &  $-0.0947 - 0.0663 i$   \\ \hline \hline
   $ t_{11,(-2,1)} = t_{11,(1,-2)} = t_{11,(1,1)} = t_{33,(-2, 1)}^* = t_{33, (1,-2)}^* = t_{33,(1,1)}^*$ & $ 0.0131 -0.0914 i$ \\ \hline
   $ t_{12,(-2,1)} = \epsilon t_{12,(1, -2)} = \epsilon^* t_{12, (1, 1)} = t_{34,(-2,1)} = \epsilon^* t_{34,(1, -2)} = \epsilon t_{34, (1, 1)}$ & $0.0706 -0.0004 i$  \\ \hline  \hline
   $ t_{11,(2, 0)} = t_{11, (-2, 2)} = t_{11,(0, -2)} = t_{33, (2,0)} = t_{33, (-2, 2)} = t_{33,(0, -2)} $ & $ -0.0005 - 0.0182 i$ \\ \hline
   $ t_{12, (2, 0)} = \epsilon t_{12, (-2, 2)} = \epsilon^* t_{12, (0, -2)} = t_{34, (-2, 2)} = \epsilon^* t_{34, (0, -2)} = \epsilon t_{34, (2, 0)}$    &   $-0.0181 + 0.0081 i$  \\  \hline \hline
   $ t_{11, (-3, 1)} = t_{11, (2, -3)} = t_{11, (1, 2)} = t_{33, (3, -2)} = t_{33, (-1, 3)} = t_{33, (-2, -1)} $   &   $0.0302 -0.0057 i$ \\ \hline
   $ t_{12, (-3, 1)} = \epsilon t_{12, (2, -3)} = \epsilon^* t_{12, (1, 2)} = t_{34, (3, -2)} = \epsilon^* t_{34, (-1, 3)} = \epsilon t_{34, (-2, -1)}$  &   $ 0.0013 + 0.0139 i  $  \\ \hline \hline
   $ t_{11, (-3, 2)} = t_{11, (1, -3)} = t_{11, (2, 1)} = t_{33, (3, -1)} = t_{33, (-2, 3)} = t_{33, (-1, -2)} $ &   $-0.0237 - 0.0097 i$   \\ \hline
   $ t_{12, (-3, 2)} = \epsilon t_{12, (1, -3)} = \epsilon^* t_{12, (2, 1)} = t_{34, (3, -1)} = \epsilon^* t_{34, (-2, 3)} = \epsilon t_{34, (-1, -2)}$   &    $ 0.0016 -0.0018 i  $ \\ \hline \hline
    $ t_{11, (3, 0)} = t_{11, (-3, 3)} = t_{11, (0, -3)} = t_{33, (-3, 3)} = t_{33, (0, -3)} = t_{33, (3, 0)} $  &  $ 0.0033 - 0.0033 i $  \\ \hline
    $ t_{12, (3, 0)} = \epsilon t_{12, (-3, 3)} = \epsilon^* t_{12, (0, -3)} = t_{34, (-3, 3)} = \epsilon^* t_{34, (0, -3)} = \epsilon t_{34, (3, 0)} $ &    $ -0.0007 + 0.0073 i$  \\ \hline
\end{tabular}
 \caption{The hopping constants between $w_{1,2}$. All the numbers are in the units of meV.}
\label{TabS:IntraHoppingWannier90}
\end{table}

\begin{figure*}[htbp]
\centering
\subfigure[\label{FigS:EneBand:2L}]{\includegraphics[width=0.22\columnwidth]{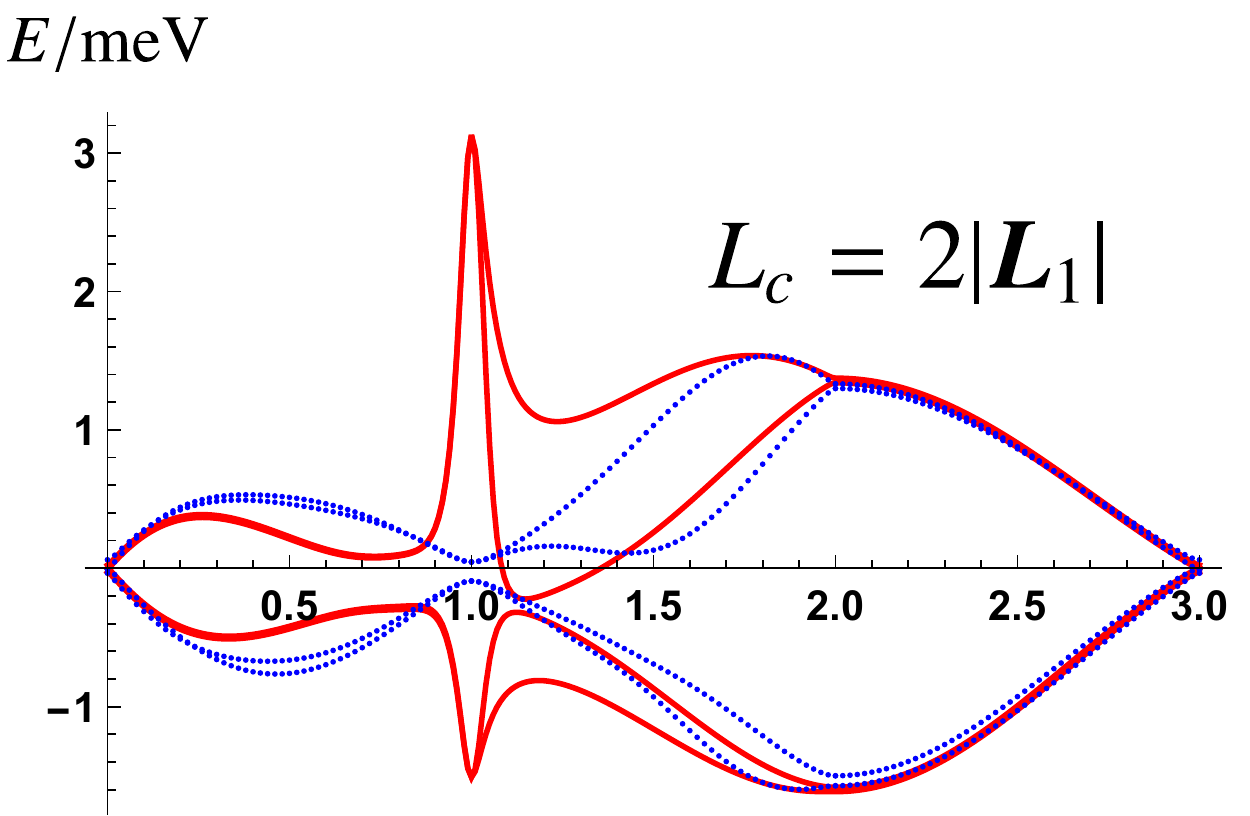}}
\subfigure[\label{FigS:EneBand:4L}]{\includegraphics[width=0.22\columnwidth]{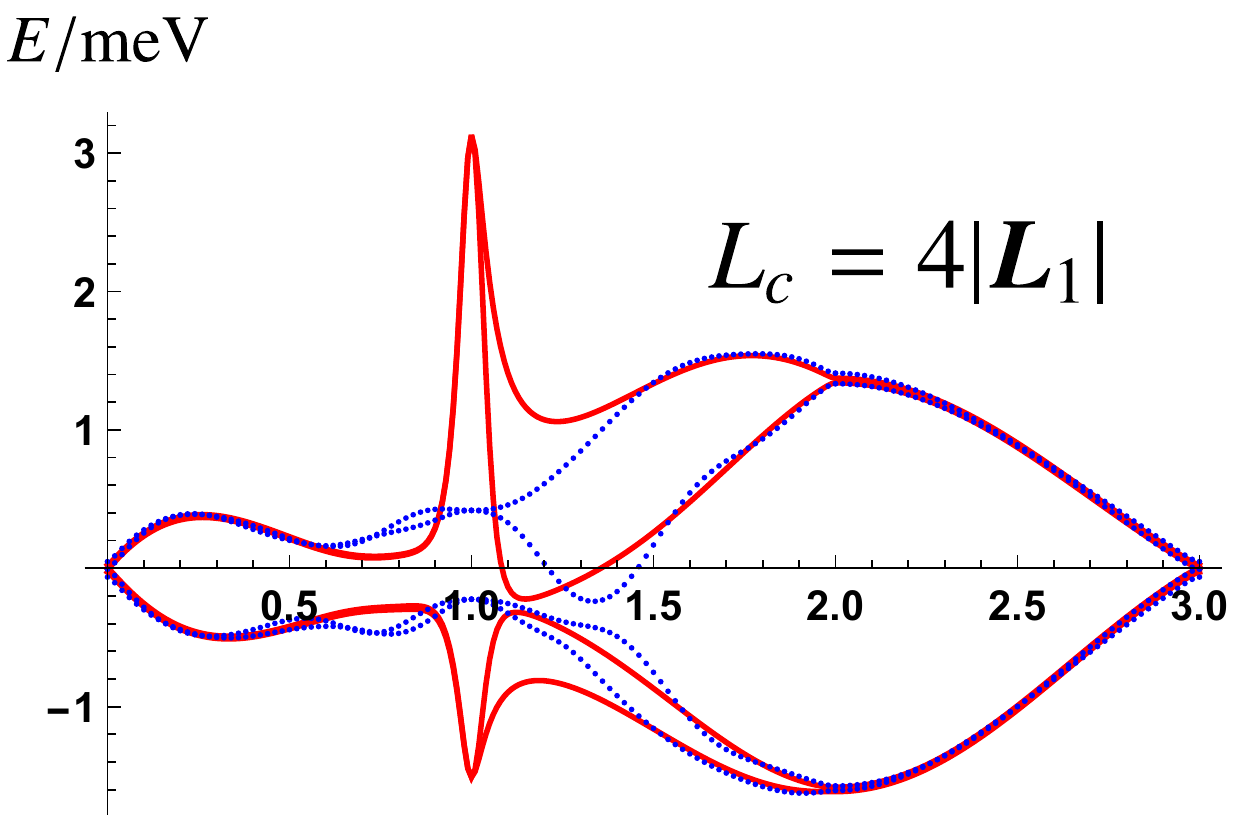}}
\subfigure[\label{FigS:EneBand:6L}]{\includegraphics[width=0.22\columnwidth]{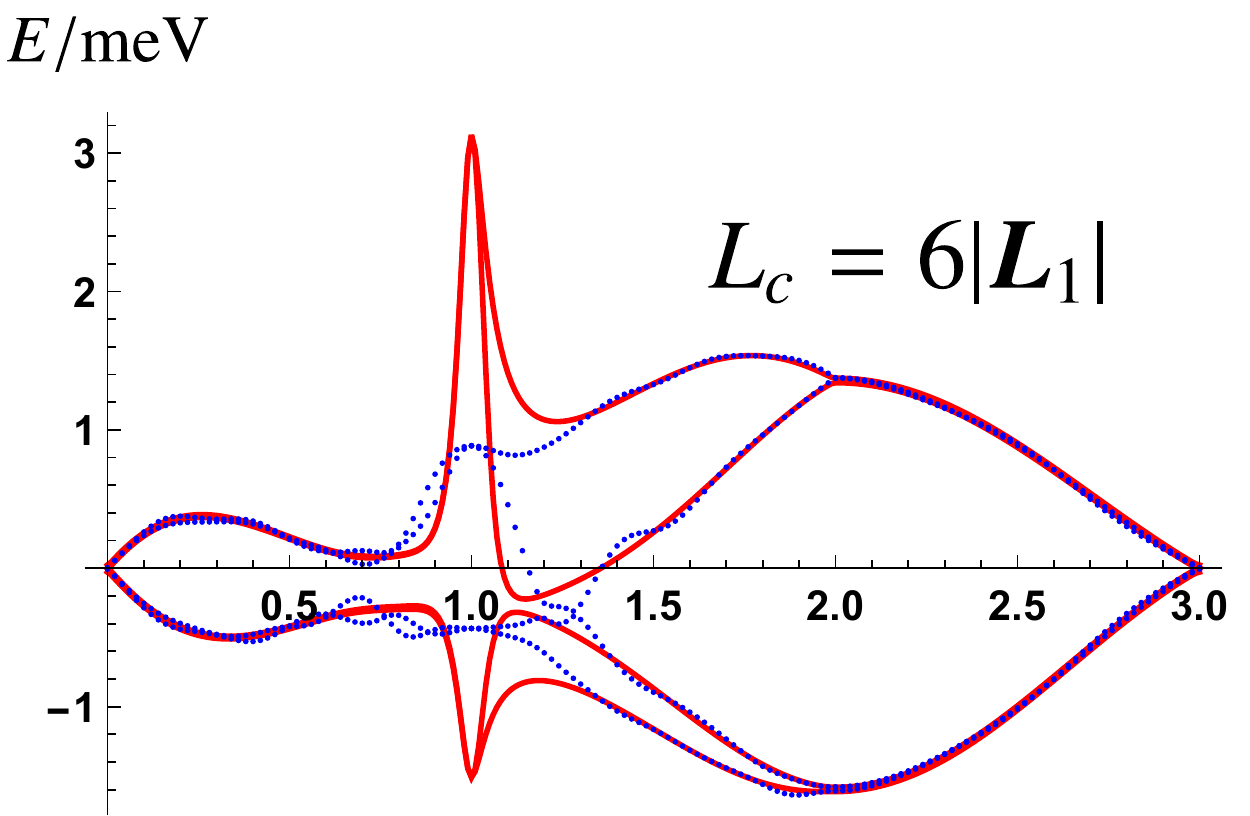}}
\subfigure[\label{FigS:EneBand:8L}]{\includegraphics[width=0.22\columnwidth]{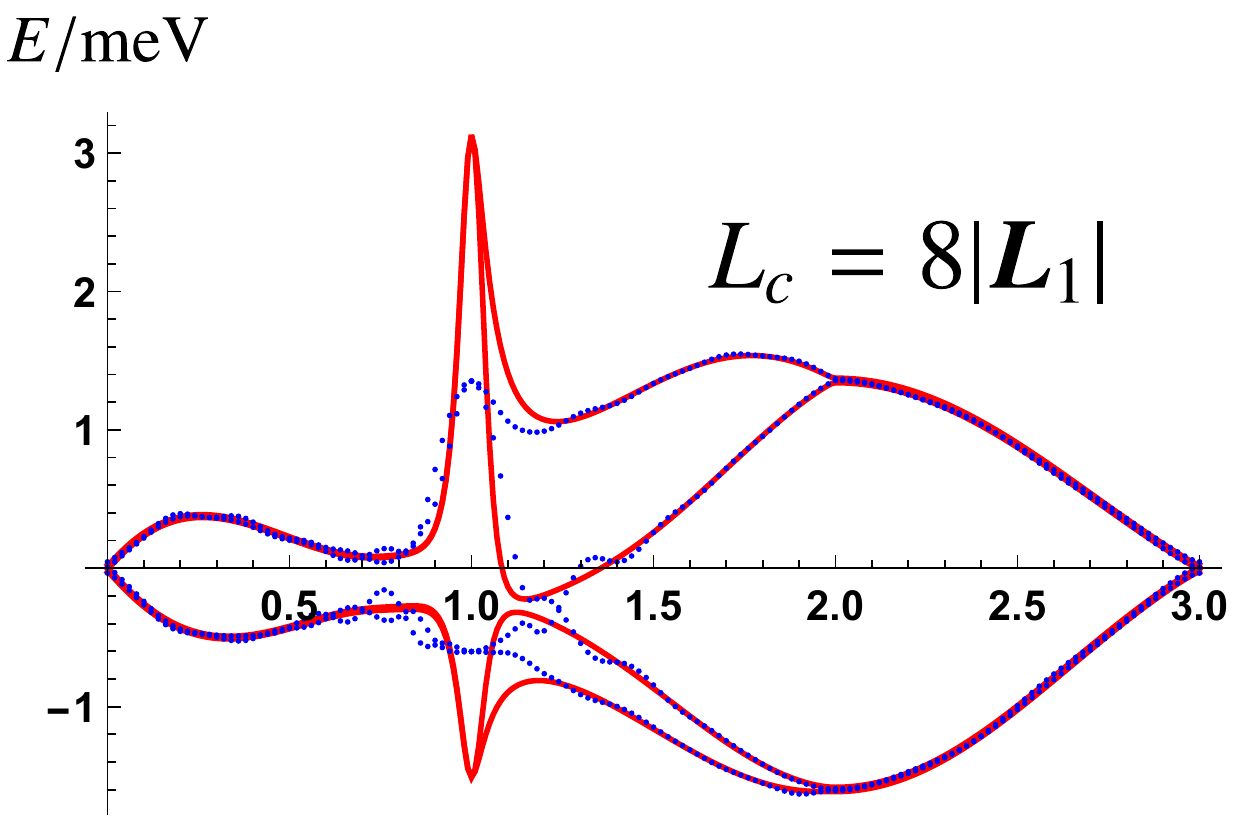}}
\caption{Comparison of the narrow band structure produced by the model given by Ref.~\cite{Koshino1} (red solid line) and the tight binding model based on the WSs (blue dots) with the range of hopping (a)  $L_c = 2L$, (b) $L_c = 4L$, (c) $L_c = 6 L$, and (d) $L_c = 8L$. }
\label{FigS:EneBand}
\end{figure*}

Fig.~\ref{FigS:EneBand} illustrates the comparison of the narrow bands produced by the model in Ref.~\cite{Koshino1} and the tight binding model based on the localized WSs with different hopping range $L_c$. For small $L_c$, most features of the band structure can be reproduced by our tight binding model, except the peaks and troughs around $\fvec \Gamma$.

\end{document}